\documentclass[10pt,journal]{IEEEtran}
\usepackage{amsmath,amsfonts,amssymb}
\usepackage[noadjust]{cite}
\usepackage{bm}
\usepackage{subcaption}
\usepackage{graphics}
\usepackage{graphicx}
\usepackage{xcolor}
\usepackage{soul}
\usepackage{hyperref}
\usepackage[absolute,showboxes]{textpos}

\TPMargin{5pt}

\newcommand{\copyrightstatement}{
    \begin{textblock}{0.84}(0.06,0.01)    
         \noindent
         \footnotesize
         \copyright  2024 IEEE.  Personal use of this material is permitted.  Permission from IEEE must be obtained for all other uses, in any current or future media, including reprinting/republishing this material for advertising or promotional purposes, creating new collective works, for resale or redistribution to servers or lists, or reuse of any copyrighted component of this work in other works. Citation information: \url{https://doi.org/10.1109/LES.2024.3472709}.
    \end{textblock}
}

\begin{document}
\copyrightstatement
\bstctlcite{IEEEexample:BSTcontrol}
\title{Using Intermittent Chaotic Clocks to Secure Cryptographic Chips}


\author{Abdollah~Masoud~Darya,~\IEEEmembership{Graduate~Student~Member,}
        Sohaib~Majzoub,~\IEEEmembership{Senior~Member,}
        Ali~A.~El-Moursy,~\IEEEmembership{Senior~Member,}
        Mohamed~Wed~Eladham,
        Khalid~Javeed,~\IEEEmembership{Senior~Member,}
        and~Ahmed~S.~Elwakil,~\IEEEmembership{Senior~Member}
\thanks{Abdollah~Masoud~Darya is with SAASST, University of Sharjah, Sharjah, UAE (abdollah.masoud@ieee.org).}%
\thanks{Sohaib~Majzoub is with the Electrical \& Computer Engineering Departments, University of Sharjah, Sharjah, UAE (smajzoub@sharjah.ac.ae).}%
\thanks{Ali~A.~El-Moursy, Mohamed~Wed~Eladham, and Khalid~Javeed are with the Computer Engineering Department, University of Sharjah, Sharjah, UAE (aelmoursy@sharjah.ac.ae, mohamed.eladham@sharjah.ac.ae, kjaveed@sharjah.ac.ae).}%
\thanks{Ahmed~S.~Elwakil is with the Electrical Engineering Department, University of Sharjah, Sharjah, UAE and the Department of Electrical \& Software Engineering, University of Calgary, Alberta, Canada (ahmed.elwakil@ucalgary.ca)}}


\maketitle

\begin{abstract}
This letter proposes using intermittent chaotic clocks, generated from chaotic maps, to drive cryptographic chips running the Advanced Encryption Standard as a countermeasure against Correlation Power Analysis attacks. Five different chaotic maps---namely: the Logistic map, the Bernoulli shift map, the Henon map, the Tent map, and the Ikeda map---are used in this work to generate chaotic clocks. The performance of these chaotic clocks is evaluated in terms of timing overhead and the resilience of the driven chip against Correlation Power Analysis attacks. All proposed chaotic clocking schemes successfully protect the driven chip against attacks, with the clocks produced by the optimized Ikeda, Henon, and Logistic maps achieving the lowest timing overhead. These optimized maps, due to their intermittent chaotic behavior, exhibit lower timing overhead compared to previous work. Notably, the chaotic clock generated by the optimized Ikeda map approaches the theoretical limit of timing overhead, i.e., half the execution time of a reference periodic clock.
\end{abstract}

\begin{IEEEkeywords}
Chaos, Cryptography, CPA.
\end{IEEEkeywords}

\IEEEpeerreviewmaketitle

\section{Introduction}

\IEEEPARstart{T}{he} emergence of the Internet-of-Things enabled the proliferation of ultra-low-power embedded systems that rely on the Advanced Encryption Standard (AES) for security \cite{yang2017hardware}. Yet, side channel attacks such as Correlation Power Analysis (CPA) threaten these embedded devices that possess limited computational capabilities \cite{padmakumar2024boosting}. CPA uses the power traces of a chip running AES to conduct a statistical analysis that can retrieve the AES key. The success of CPA attacks relies on synchronizing the collected power traces, which is possible with devices driven by periodic clocks. As a countermeasure against CPA attacks, techniques such as Random Dynamic Frequency Scaling (RDFS) introduce misalignment in the power traces, thus preventing CPA from extracting the AES key \cite{dao2021correlation}. However, generating and using random clocks adds significant hardware and performance costs.\par 

Recently, chaotic clocking was shown to be a simpler alternative to RDFS, using autonomous chaotic oscillators \cite{el2020chaotic}. While these autonomous systems protected against CPA attacks, they imposed considerable timing overhead compared to conventional periodic clocking. To address this issue,  \cite{el2022using} proposed using non-autonomous chaotic oscillators to generate chaotic clocks, thus reducing the timing overhead. However, the main drawback of non-autonomous chaotic oscillators is that they must be driven by an external periodic clock, the frequency of which affects the dynamic behavior of the system---i.e., it is not chaotic at all values of this clock frequency.\par

Chaotic maps are free of the limitation of non-autonomous chaotic oscillators since the periodic clock is used for iterating the system but its frequency does not influence its dynamics. Although chaotic maps have been used previously for hardware security in embedded systems \cite{harsha2017novel}, their application in driving cryptographic chips has not been explored. However, \cite{xiao2021countermeasure} used a hybrid chaotic map to generate random numbers that mask the intermediate data in AES. Also, \cite{tran2023dynamic} used a Gold-code-based chaotic clock, which combines two different types of random number generators so that each AES encryption is conducted using a random Gold sequence. \par

This work proposes driving AES chips with chaotic clocks generated from chaotic maps. Specifically, five well-known maps, three of which are 1-D and two are 2-D, are investigated in this study. An exhaustive search is conducted for all five maps to find their optimal parameters that minimize the timing overhead of the resulting chaotic clock. The results show that while all maps successfully protect against CPA attacks, the highest throughput is achieved by the Ikeda map. Further investigation reveals that the optimized maps exhibit \emph{intermittent chaos}. Experimental results validate these findings.\par

\section{Chaotic Maps}\label{Background}
Chaotic maps are discrete dynamical systems that produce chaotic values sensitive to initial conditions.  Many 1-D and 2-D chaotic maps have been proposed and studied \cite{naik2022review}. The five  maps investigated in this work are:

\subsection{The Logistic Map}
 This 1-D map is represented by the equation \cite{lawrance2003binary}
\begin{equation}
x[n+1]=\lambda \cdot x[n] \cdot \left( 1- x[n]\right),
\end{equation}
where $\lambda$ is the tuning parameter, $x[n]\in \left(0,1\right)$, and $x[0]$ is initialized to $0.1$. To generate a chaotic clock, $x[n]$ is discretized as
\begin{equation}\label{eq2}
\text{CLK}_\text{Logistic}[n]=
\begin{cases}
     0, & \text{if } x[n]< c,\\
     1, & \text{otherwise},
\end{cases}
\end{equation}
where we select $c=0.5$ based on the results  in \cite{kanso2009logistic}.\par

\subsection{The Bernoulli Shift Map}
This is another 1-D map, represented by the equation \cite{sukegawa2022perturb}
\begin{equation}
x[n+1]=2 \cdot x[n]~(\text{mod}~1-\epsilon),
\end{equation}
where $\epsilon$ is the tuning parameter. Similar to the Logistic map, $x[0]$ is initialized to $0.1$ and the chaotic clock $\text{CLK}_\text{Bernoulli}[n]$ is generated using \eqref{eq2} with $c=0.5$.\par

\subsection{The Tent Map}
This map is defined by the equation \cite{crampin1994chaotic}
\begin{equation}
x[n+1]=
\begin{cases}
     2 \cdot k \cdot x[n], & 0\leq x < 0.5,\\
     2 \cdot k \cdot (1-x[n]), & 0.5\leq x < 1,
\end{cases}
\end{equation}
where $k$ is the tuning parameter. Similar to the Logistic and Bernoulli shift maps, $x[0]$ is initialized to $0.1$, and the chaotic clock $\text{CLK}_\text{Tent}[n]$ is obtained from \eqref{eq2} with $c=0.5$.\par

\subsection{The Henon Map}
This is a 2-D map given by the equations \cite{henon1976two}
\begin{equation}
\begin{split}
x[n+1]&=y[n] + 1 - a \cdot x[n]^2,\\
y[n+1]&=b \cdot x[n],
\end{split}
\end{equation}
where $a$ is the tuning parameter and $b=0.3$. Two different chaotic clocks can be obtained from this map using a \emph{signum} function: $\text{CLK}_\text{Henon-x}[n] = sgn(x[n])$ and $\text{CLK}_\text{Henon-y}[n] = sgn(y[n])$, respectively. This work uses $\text{CLK}_\text{Henon-x}[n]$ due to the similar frequencies of the two clocks..\par

\subsection{The Ikeda Map}
The 2-D Ikeda map is described by \cite{galias2002rigorous}
\begin{equation}
\begin{split}
&x[n+1]=p+B \cdot (x[n]  \cdot \cos t - y[n]  \cdot \sin t),\\
&y[n+1]=B \cdot (x[n]  \cdot \sin t - y[n]  \cdot \cos t),
\end{split}
\end{equation}
where $B$ is the tuning parameter, $t=0.4 - 6/(1+x^2+y^2)$, $p=1$, and $x[0]$ and $y[0]$ are initialized to $0$. The chaotic clocks are generated as $\text{CLK}_\text{Ikeda-x}[n] = sgn(x[n])$ and $\text{CLK}_\text{Ikeda-y}[n] = sgn(y[n])$, respectively. In this work, $\text{CLK}_\text{Ikeda-y}[n]$ was used because it is faster than $\text{CLK}_\text{Ikeda-x}[n]$.\par

\begin{figure}[!t]
\centering
\includegraphics[width=1\columnwidth]{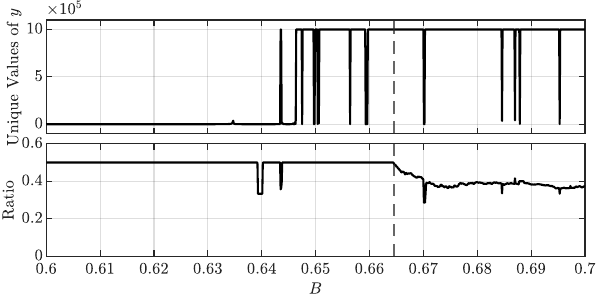}
\caption{\small{Search of optimal $B$ which (i) maintains chaotic behavior (upper trace) and (ii) maximizes the chaotic clock ratio (lower trace). The optimal $B$ of the Ikeda map is represented by the dashed line.}}
\label{fig1}
\end{figure}

\section{Results and Discussion}\label{RD}
This section discusses the optimization of the chaotic maps to reduce timing overhead. The optimized maps are then used to secure an FPGA running AES-128 against CPA attacks. The timing, power, and area overheads are experimentally evaluated for all five chaotic maps.\par

\begin{table}[!t]
\caption{\small{Tuning results of the chaotic maps and comparison with related work}}\label{tab1}
\centering
\resizebox{\columnwidth}{!}{%
\begin{tabular}{|l|c|c|}
\hline
\textbf{System} & \textbf{\begin{tabular}[c]{@{}c@{}}Parameter Range\\ $\{$Optimum Value$\}$\end{tabular}} & \textbf{Output Clock Ratio} \\ \hline
Bernoulli Shift Map & $\epsilon \in(0,0.5) \{0.0005\}$ & $0.2502$ \\ \hline
Tent Map & $k \in(0.5,1] \{0.6064\}$ & $0.3867$ \\ \hline
Logistic Map & $\lambda \in(3.5,4) \{3.5895\}$ & $0.4213$ \\ \hline
Henon Map---x & $a \in(1,1.5) \{1.0909\}$ & $0.4286$ \\ \hline
Ikeda Map---y & $B \in[0.6,0.9] \{0.6645\}$ & $0.4963$ \\ \hline
SSJCO \cite{el2020chaotic} & - & $0.0014$ \\ \hline
TWCO \cite{el2020chaotic} & - & $0.0006$ \\ \hline
Lorenz \cite{el2022using} & - & $0.0031$ \\ \hline
NCS$_{P=8}$ \cite{el2022using} & - & $0.1197$ \\ \hline
RDFS \cite{dao2021correlation} & - & $0.2976$ \\ \hline
Gold Code \cite{tran2023dynamic} & - & $0.2497$\footnotemark \\ \hline
\end{tabular}%
}
\end{table}

\footnotetext{This value was calculated based on our reproduction of the work in \cite{tran2023dynamic}.}

\subsection{Parameter Tuning}\label{optm}
 An exhaustive search was conducted using MATLAB to find the optimal tuning parameter for each map. First, the chaotic nature of each map was confirmed by taking $10^6$ samples from the $x[n]$ or $y[n]$ values and ensuring they were unique and non-repeated. Second, the parameter was tuned to maximize the ratio of the generated chaotic clock frequency (per $10^6$ samples) to the frequency of the periodic clock used to iterate the maps (reference periodic clock). An example of these two steps is presented in Fig. \ref{fig1} for the Ikeda map and the results of the exhaustive search are listed in Table \ref{tab1}. The output clock ratios in Table \ref{tab1}, for each system, are found by dividing the number of output clock cycles by the reference periodic clock cycles. The frequency of the clocks produced by the chaotic maps must be $\leq0.5\times$ the reference clock frequency, i.e., it takes at least two cycles for the reference periodic clock to generate a single chaotic clock cycle. Therefore, the closer the calculated ratio is to $0.5$, the faster the chaotic system is, and hence, the lower the timing overhead.\par
It is seen from Table \ref{tab1} that the highest output chaotic clock ratio was achieved by the Ikeda map, closely followed by the Henon and Logistic maps. On the other hand, the Bernoulli shift map was the slowest of the five maps. Compared to the proposed maps, the autonomous and non-autonomous chaotic systems proposed in \cite{el2020chaotic} and \cite{el2022using} have significantly higher timing overheads. Meanwhile, the RDFS-based clock proposed in \cite{dao2021correlation} and the Gold-code-based chaotic clock from \cite{tran2023dynamic} exhibit comparable, yet still higher, overheads.\par

\begin{figure}[!t]
\centering
\includegraphics[width=1\columnwidth]{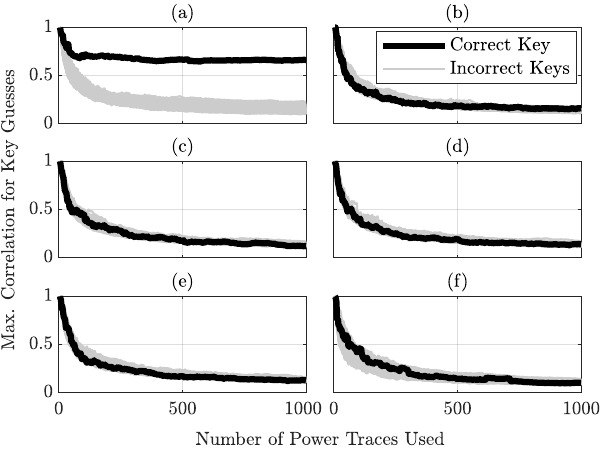}%
\caption{\small{Results of the CPA attacks on the AES chip driven by different clocks:  (a) unprotected periodic clock, (b)  Logistic map chaotic clock, (c) Bernoulli shift map chaotic clock, (d) Henon map chaotic clock, (e) Tent map chaotic clock, and (f) Ikeda map chaotic clock.}}
\label{fig_sim}
\end{figure}

\subsection{CPA Analysis}
The openSCA MATLAB toolkit \cite{mangard2008power} was used to test the resilience of an FPGA chip, driven by chaotic clocks from the optimized chaotic maps in Section \ref{optm}, against CPA attacks. We assume that the device under attack maintains the same AES key for at least $1\,000$ AES runs. Therefore, using $1\,000$ experimental power traces of $25\,000$ samples each (provided by openSCA), corresponding to the implementation of the first AES round, CPA was conducted following the same procedure described in \cite{el2020chaotic}. This includes: a) generating the chaotic clock cycles in MATLAB, b) using them to scale the experimental openSCA power traces by resampling the power traces to match the frequency of the chaotic clock cycles, and c) implementing CPA using the openSCA toolkit.\par
As shown in Figs. \ref{fig_sim}(b)--(f), all five chaotic maps successfully protected against CPA attacks, as the key could not be recovered even with $1\,000$ power traces. However, CPA deduced the key of the unprotected AES, i.e., running on a periodic clock, using $17$ traces (see Fig. \ref{fig_sim}(a)).\par

\begin{figure}[t]
\centering
\includegraphics[width=1\columnwidth]{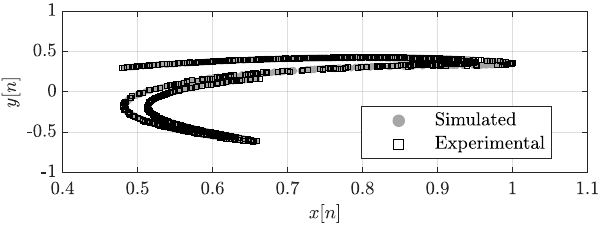}
\caption{\small{Experimental Ikeda map ($B=0.6656$) compared to the simulated Ikeda map ($B=0.6645$) for $3\times10^4$ samples.}}
\label{fig:setup}
\end{figure}%

\subsection{Experimental Setup}
Two FPGAs were used to experimentally test the proposed chaotic maps: an Altera DE2 and a tinyFPGA BX. The Altera DE2 is used to execute the chaotic maps and produce a chaotic clock that is then supplied to the tinyFPGA BX running AES-128. The tinyFPGA was chosen to emulate a low-power cryptographic chip running AES-128.\par
To measure the power consumption of the tinyFPGA BX while running AES-128, a Tektronix CT2 current probe was connected to a Digilent Analog Discovery 2 board. The mean power consumption was calculated by averaging the power values over $15$ AES-128 runs. The mean frequency values of the DE2 chaotic clock output were found by taking $10^6$ samples from the DE2 board and then calculating the frequency offline to ensure accuracy. Additionally, using $50$ AES-128 runs with the same key and input text, the mean number of cycles to execute AES by the tinyFPGA BX was found to be $504.69$ cycles.\par
All five chaotic maps were implemented using the Altera DE2 board running at $7.14$ MHz (reference periodic clock). This frequency was chosen to increase the resolution of the timing/power observations. Since the Ikeda map contains the trigonometric functions \emph{sine} and \emph{cosine}, the CORDIC algorithm \cite{garrido2015cordic} was used for efficient implementation on the DE2 board. Consequently, due to the approximations of the CORDIC algorithm, the optimal value of $B\!=\!0.6645$, found previously in simulations (see Table \ref{tab1}), was changed to $B\!=\!0.6656$ which reduced the simulation-based chaotic clock ratio from $0.4963$ to $0.4588$.  The negligible effect of this change of $B$ on the Ikeda chaotic attractor can be seen in Fig. \ref{fig:setup}. The optimal values for the other maps remained unchanged.\par

\begin{table}[!t]
\caption{\small{Experimental timing, power, and area overheads}}\label{tab2}
\centering
\resizebox{\columnwidth}{!}{%
\begin{tabular}{|l|c|c|c|c|c|c|}
\hline
\textbf{Metric}&\multicolumn{1}{c|}{Periodic}&\multicolumn{1}{c|}{Bern.}&\multicolumn{1}{c|}{Tent}&\multicolumn{1}{c|}{Log.}&\multicolumn{1}{c|}{Henon}&\multicolumn{1}{c|}{Ikeda} \\ \hline
\begin{tabular}[c]{@{}l@{}}Mean Freq.\\(MHz)\end{tabular} &7.14&1.78&2.81&2.97&3.06&3.20\\ \hline
\begin{tabular}[c]{@{}l@{}}Mean Exec.\\Time ($\mu$s)\end{tabular} &70.68&283.53&179.61&169.93&164.93&157.72\\ \hline
\begin{tabular}[c]{@{}l@{}}Mean Power\\(mW)\end{tabular} &4.99&3.07&3.47&3.62&3.56&3.46\\ \hline
\begin{tabular}[c]{@{}l@{}}Design Area\\(Logic Elem.)\end{tabular} &-&2.91\%&1.04\%&1.54\%&1.51\%&34.83\%\\ \hline
\end{tabular}%
}
\end{table}

The performance results in terms of the mean frequency, mean AES execution time, mean power consumption, and utilized logic of the Altera DE2 are presented in Table \ref{tab2}. The experimental chaotic clock ratio of the chaotic maps is found by dividing the mean frequency of the chaotic clock of each map by the reference periodic clock. Using the values of Table \ref{tab2}, the experimental chaotic clock ratios are $0.25$, $0.39$, $0.42$, $0.43$, and $0.45$ for the Bernoulli Shift, Tent, Logistic, Henon, and Ikeda maps, respectively. These experimentally obtained ratios agree well with the simulated ratios from Table \ref{tab1}.\par
The power consumption of the chaotic-clock-driven tinyFPGA BX running AES-128 is $62\%$, $70\%$, $73\%$, $71\%$, and $69\%$ that of the periodic clock-driven FPGA using the Bernoulli Shift, Tent, Logistic, Henon, and Ikeda maps, respectively. While the Ikeda map produces the fastest clock, it is more complex due to the presence of trigonometric functions in its formulation. Thus, it utilized $34.8\%$ of the total logic elements compared to $1.5\%$ for the Henon and Logistic maps. Therefore, the Henon and Logistic maps are preferable when computational resources are limited.\par

\begin{figure}[!t]
\centering
\includegraphics[width=1\columnwidth]{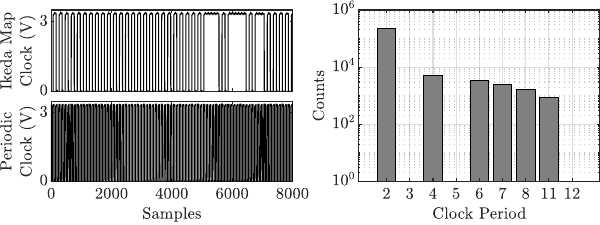}
\caption{\small{Samples of the experimental chaotic clock of the Ikeda map (top left trace) and a periodic clock (lower left trace) from the DE2 board. Right plot: statistical distribution of the experimental clock periods of the Ikeda map using $10^6$ samples from the DE2.}}
\label{fig:samp}
\end{figure}%

\subsection{Discussion}
It is important to understand the reason behind the high speed obtained from the optimized maps compared to related work. To illustrate this, a sample chaotic clock from the Ikeda map is plotted in Fig. \ref{fig:samp}. From this figure, it is immediately recognized that these chaotic clocks exhibit \emph{intermittent behavior} which is characterized by \emph{long} intervals of periodicity interrupted by sudden \emph{bursts} of chaos \cite{elaskar2023review}. This behavior explains the improved timing performance of the optimized maps, as the long periodic intervals result in an average frequency nearly equal to half the frequency of the reference periodic clock operating the DE2 (see lower left trace of Fig. \ref{fig:samp}). To confirm this, the histogram in Fig. \ref{fig:samp} shows the distribution of clock period lengths of the Ikeda map where it is evident that most periods have a length of $2\times$ the period of the reference clock.\par
Intermittency is a complex dynamical behavior with more than ten different types, as discussed in detail in \cite{elaskar2023review}. It is clear from our results that no value of $k$ can make the Tent map outperform the Ikeda map operating at its optimal $B$ value despite the simplicity of the Tent map when compared to the complexity of the Ikeda map. This means that choosing the type of map, and hence the corresponding type of intermittent behavior, is critical. This, however, requires another separate comprehensive study. Nevertheless, to assert that intermittent chaos can provide speed while maintaining security, we used the intermittent Lorenz system described in \cite{inter-Lorenz} with $\sigma\!=\!10,\eta\!=\!8/3,\rho\!=\!166.25$ and compared it with the performance of the same system with $\sigma\!=\!10,\eta\!=\!8/3,\rho\!=\!28$ from \cite{el2022using}, both with a chaotic clock generated as $sgn\left(x\left(t\right)\right)$. We found that the intermittent Lorenz system was $5.8\times$ faster, thus improving the mean chaotic clock frequency from $4.54$ kHz to $26.5$ kHz using a reference periodic clock of $7.14$ MHz. Despite this speed improvement when using intermittent clocking of the Lorenz system, it is clear that autonomous chaotic systems still cannot outperform any of the chaotic maps in this work.\par

\section{Conclusion}\label{Conc}
This work proposes protecting AES-running cryptographic chips using chaotic clocks generated from chaotic maps. In \cite{el2020chaotic}, we verified that chaotic clocks generated by autonomous chaotic oscillators can protect against CPA attacks, but at the cost of high timing overhead. In \cite{el2022using}, we demonstrated that non-autonomous chaotic oscillators can provide security with comparatively lower overhead, though they are sensitive to the frequency of the periodic clocks driving them. This work shows that clocks derived from optimized chaotic maps can protect against CPA attacks with superior timing performance due to their intermittent chaotic behavior. However, since intermittency has various types, our future work will focus on identifying the optimal type of intermittency and the corresponding maps. Additionally, we will explore the performance of the proposed chaotic maps on other cryptographic cores, such as Elliptic Curve Cryptography and Secure Hash Algorithms.\par

\section*{Acknowledgments}
The authors would like to thank Oruba Alfawaz for her assistance in the FPGA implementation of the Tent map.

\bibliographystyle{IEEEtran}
\bibliography{Manuscript}

\end{document}